\documentclass{revtex4}

\usepackage{amsfonts}
\usepackage{amsmath}
\usepackage[latin2]{inputenc}
\usepackage{graphicx}
\def\i1{\mathcal 1}

\def\1{\mathrm 1}
\def\bbC{{\mathbb C}}

\def\x{\otimes}
\def\>{\rangle}
\def\<{\langle}
\def\Tr{\mathrm Tr}

\begin{document}
\title{\textbf{Exploring phase transitions by finite-entanglement scaling of MPS in the 1D ANNNI model}}
\author{Adam Nagy}

\affiliation{Dipartimento di Fisica Teorica, Universit\`a di Trieste
\\Strada Costiera 11,
34014 Trieste, Italy and \\
Department of Mathematical Analysis, BUTE
H-1521 Budapest, POB 91, Hungary
}

\date{\today}

\begin{abstract}
We use the finite-entanglement scaling of iMPS to explore supposedly infinite order transitions.
This universal method may have lower computational costs than finite-size scaling. 
To this end we study possible MPS based algorithms to find the ground states of the transverse ANNNI model in a spin chain with first and second neighbour interactions and frustration. The ground state has four distinct phases with transitions of second order and one of supposedly infinite order, Kosterlitz-Thouless transition. To explore phase transitions in the model, we study general quantities like the correlation length, entanglement entropy and the second derivative of the energy with respect to the external field, and test the finite-entanglement scaling. We propose a scaling ansatz for the correlation length of a non critical system in order to explore
infinite order transitions.  
This method provides considerable less computational costs compared to the finite-size scaling method in Ref. \cite{annni} and
quantities obtained by applying fixed boundary conditions (like domain wall energy in Ref. \cite{annni}) are omitted. 
The results show good agreement with previous studies of finite-size scaling using DMRG.
\end{abstract}


\maketitle

\section{Introduction}

Recently, numerical simulations of many-body quantum systems based on matrix product states (MPS) and their generalizations have called increasing attention as they offer an efficient way to study properties of their ground states and even thermal states. One of the main reasons of their popularity is  their applicability to--in principle--any Hamiltonians describing only local interactions on a spin chain and their possible generalizations to higher dimensions. The use of the so-called projected entangled pair states (PEPS) to approximate ground states of such kind of Hamiltonians defined in higher dimensions (e.g. on a square lattice) is considered very promising. (See e.g. Ref.\cite{review} and references therein for a review of the topic.)
These methods can provide efficient simulation of systems when other methods such as quantum Monte Carlo break down, e.g. in case of certain frustrated systems. It means for example that one can circumvent the so-called sign problem present in the Monte Carlo simulation.  (Note that there exist systems for which there is no other method available, not considering exact diagonalization which works only for small system sizes. One such system is e.g. the frustrated XX model. Ref. \cite{sts})

Obviously one of the main goals of investigating ground states is to explore phase transitions and characterize them as precisely as possible. Despite the fact that by definition, the MPS can only describe exponentially decaying correlations, one can obtain accurate results for certain quantities of phase transitions like in the Ising model as it has been demonstrated in nearly all of the papers presenting an MPS based algorithm. (See e.g. Ref. \cite{farhi} where it has also been used for an Ising model defined on a binary tree.)
Models with not only nearest neighbor (NN) interactions and higher than second order transitions have not been studied extensively by means of MPS.

This paper has two goals: Find a suitable efficient algorithm based on MPS to simulate Hamiltonians with second neighbor interactions and frustration,
as well as demonstrate the use of universal tools, especially finite-entanglement scaling to study different types of phase transitions.
Its validity arises from general facts about critical points and the definition of the MPS, hence it can be applied quite universally.
As opposed to the usual finite-size scaling it may also have a better scaling for the computational costs if using the same maximal bond dimensions for both cases (see below).  
Here the 1-dimensional version of the transverse axial next nearest neighbor Ising (ANNNI) model has been chosen which is non-integrable and possesses a quite complex phase diagram, despite being the simplest model with NNN interactions.
This model is notable due to several facts. As earlier studies have shown, unlike the Ising model, it has a critical region, i.e. a 2D parameter region where it is critical. A probably infinite order phase transition also takes place (besides second order phase transitions). As mentioned earlier, the detection of an infinite (e.g. Kosterlitz-Thouless) transition is a notoriously hard computational challenge, that's why a computationally beneficial method is needed. The same holds for exploring the physics of a critical region.

In order to study such a model one can choose a finite chain of length $L$ with periodic (PBC) or open boundary conditions (OBC) and then look at the scaling of some physical quantities depending on $L$ and the bond dimension $D$. It is known that the overall computational costs scales $O(D^3)$ in the best case. The latter happens for example when the correlations are very small $\xi\ll L$ as it is discussed in Ref. \cite{ujpirvu}, however when $\xi\approx L$ e.g. it can be as large as  $O(D^5)$ and the precise behaviour of the scaling as a function of $D$ and $L$ is unclear.

Note that in this paper, iMPS and imaginary time evolution are used, hence one works directly on the infinite lattice and the only scaling parameter is $D$.
In this case the computational cost in every small time step $\Delta t$ scales as $O(D^3)$. Roughly speaking, convergence can be reached if: $\exp(t\Delta E)=\exp(n\Delta t\Delta E)\ll 1$ with the  $\Delta E$ energy gap and $n$ the number of time steps. Approaching to a critical point, $\Delta E$ vanishes as a function of the effective correlation length $\xi_D$, and at the same time we also have a scaling relation like $\xi_D\propto D^{k}$ (see below).
With this in mind, it can be estimated how the overall computational time scales as a function of $D$ and one sees why it is slower than $O(D^3)$ near to a critical point.
For instance, if $\Delta E\propto \xi_D^{-\alpha} \propto D^{-\alpha k}$, $\alpha>0$ then $n$ should scale as $O(D^{\alpha k})$ that makes the overall scaling $O(D^{3+\alpha k})$.
Previous work about finite-D scaling can be found e.g. in \cite{pollmann} and \cite{tagliacozzo}.

Now the idea is to study the scaling of physical quantities not as a function of the system size, but simply as a function of the bond dimension $D$. Instead of finite-size scaling this can be called as 'finite-D' or 'finite-entanglement' scaling as a MPS with a given $D$ can contain a finite amount of entanglement.
As it will be argued in Section IV. the overall computation cost of finite-D scaling can be considerably reduced compared to that of the usual finite-size scaling. In fact, this turns out to be the case for the presented model as well as for the Ising and Heisenberg models, and essentially this depends on the central charge. 
The key point to this computational cost reduction is our off critical scaling ansatz (\ref{finitescaling}) for the correlation length as a function of $D$ which turns out to be computationally more beneficial than that of used in Ref. \cite{annni} for the entanglement entropy.

Clearly, our methods can be  applied straightforwardly to any other 1D model with NN and NNN interactions, such as the $J_1-J_2$ Heisenberg model, but can also be adapted for the 2D version using PEPS.
Despite the polynomial computational time scaling, unfortunately its exponent is so high in the 2D case that in practice one can apply only very small bond dimensions providing only a few and ambiguous results that are not yet sufficient for numerical analysis.

\section{The 1D transverse ANNNI model}
Consider the following Hamiltonian on an infinite  chain of spin-$1/2$ particles
$$
H=-\sum_{i}(J_1\sigma_i^{z}\sigma_{i+1}^{z}+J_2\sigma_i^{z}\sigma_{i+2}^{z}+h\sigma^x_i)=H_{Ising}+H_{NNN}.
$$
This can be considered as the 1 dimensional version of the lattice model with axial next nearest neighbor interactions (ANNNI).
If the next nearest neighbor (NNN) interaction term describes antiferromagnetic interactions ($J_2<0$) and $J_1>0$ then the 
system is frustrated.
Note that basically two parameters determines the physical properties of  model e.g. $\kappa=J_2/J_1$ and $h/J_1$ . Changing the basis of every second spin appropriately by rotating around the axis $x$ leads to $Z^{2i}\rightarrow -Z^{2i}$, so one can switch from an antiferromagnetic NN interaction to a ferromagnetic one and vice-versa, leaving the NNN interaction unchanged. However, there is no rotation switching all couplings to ferromagnetic, thus the sign of the NNN interactions is crucial and besides the magnetic field of $h/|J_1|$ the ratio $\kappa=J_2/|J_1|$ plays the key role in determining the physical properties.
From now on let us set $J_1=1>0$ and $J_2<0$ yielding competing ferro-and antiferromagnetic NN and NNN couplings.
\section{MPS states and algorithms}

The MPS representation of an infinite, translational invariant chain of $d$-dimensional systems looks like
$$
|\Psi\rangle=\sum_{s_1,s_2..=1}^d\Tr(A^{s_1}_1A^{s_2}_2..)|s_1,s_2...\rangle
$$
with site independent sets of matrices $\{A^s\}_s$ of dimension $D\times D$. The normalization constraint requires that the largest eigenvalue of the transfer matrix $E=\sum_s A^{s}\x \overline{A^{s}}$ has to be unique and equals to $1$.
If it's degenerate (as it happens e.g. for the state $\Psi=c_0\uparrow\uparrow\uparrow...+c_1\downarrow\downarrow\downarrow...$)  then it can cause troubles in some types of the algorithms that work with unique leading eigenvectors (Ref. \cite{pirvu}, Ref. \cite{orus}).
Nevertheless, in order to overcome this one can apply tricks like appropriate rotations of some basis vectors, but due to the lots of computations like leading eigenvectors, inverses and squares of matrices these algorithms turn out to be costly.
Another representation Ref. \cite{itebd} uses the Schmidt coefficients in the diagonal matrices $\lambda^{i}$ explicitly as follows
$$
|\Psi\rangle=\sum...\Gamma^{(i),s_i}\lambda^{(i)} \Gamma^{(i+1),s_{i+1}}\lambda^{(i+1)}...|s_i\rangle |s_{i+1}\rangle ...
$$
with the normalization constraints
$$
\Tr \Big( \lambda^{(i)}\Big) ^2=\1
$$ 
$$
\sum_{s=1}^d \Big( \Gamma^{(i),s_i}\Big) ^+ \Big( \lambda^{(i)}\Big) ^2 \Gamma^{(i),s_i}=\1_{D_i}\qquad
\sum_{s=1}^d \Big( \Gamma^{(i),s_i}\Big) ^+ \Big( \lambda^{(i-1)}\Big) ^2 \Gamma^{(i),s_i}=\1_{D_{i-1}}
$$
in the general case when the dimensions of $\lambda^{(i-1)}$ and $\lambda^{(i)}$ can be different.
The infinite time evolving block decimation (iTEBD) Ref. \cite{itebd} uses this description and in case of NN interactions two neighboring sites are updated after the action of $e^{-\delta tH_{i,i+1}}$. One builds up a larger matrix occupying the MPS matrices of the two sites $i$ and $i+1$ and decomposing them through a singular value decomposition to obtain the new matrices. A normalization step can be interpreted as a zero time update: we just build a larger matrix and decompose it.
The details can be found in Ref. \cite{farhi}.

Several ways exist to do the updates in the case at hand.
One option is the use of `superspins' ( made of 2 neighboring spins).
To this end let us join together every second neighboring pairs of spins to form block spins with $2\times 2=4$ dimensional basis as
$ |t_{i}, t_{i+1}\rangle |t_{i+2}, t_{i+3}\rangle=|s_{i}\rangle|s_{i+2}\rangle$ with $t_i\in\{0,1\}$ and $s_i\in\{0,1,2,3\}$. In this way we get a translational invariant Hamiltonian with only NN interactions acting between these 4 dimensional block spins so we can use the simplest algorithms designed for this case but with $d=4$ and a bit more complicated NN term.

In our scheme, for example the operator $\sigma^z_{i}\x\sigma^z_{i+1}\in M_4(\bbC)$ acting on the sites $i,i+1$ in the original lattice is interpreted as an operator acting on the single superspin labelled by $i$,
while $\1_{i}\x\sigma^z_{i+1}\x\sigma^z_{i+2}\x\1_{i+3}\in M_{16}(\bbC)$ describes an interaction between the superspins at the position $i$ and $i+2$.
Similarly $\sigma^z_{i}\x\1_{i+1}\x\sigma^z_{i+2}\x\1_{i+3}\in M_{16}(\bbC)$ is a NN interaction between the superspins,
however it describes a NNN coupling between the original spins.
With this in mind one can easily construct the operators 
$$
\exp(\epsilon_1 \sigma^z_{i} \sigma^z_{i+1}+\epsilon_2 \sigma^z_{i} \sigma^z_{i+2})
$$
in the new basis.

We make use of the `superspins' and the update method described in Ref.\cite{farhi}. The computation time in every step scales as $O(D^3)$ but as opposed to the previous method like in Ref.\cite{orus}, one does not need to compute neither leading eigenvectors nor
inverses or squares of non diagonal matrices hence it is faster and there is no problem with degenerate eigenvalues.

\section{Finite-entanglement scaling to detect phase transitions}
In order to characterize phase transitions using iMPS one needs to study how some special quantities scale as a function of the applied bond dimension $D$. Let us denote the entropy of entanglement computed from the iMPS by $S_D$ and the correlation length $\xi_D$. It is known that in the critical points $S_D$ converges logarithmically with corrections of order $O(1/\log D)$ as 
\begin{equation}
S_D=\frac{c k}{6}\log D
\label{entropyscaling}
\end{equation}
with $c$ being the central charge and $k$ only depends on  $c$ for large enough $D$  as 
\begin{equation}
k=\frac{6}{c\left( \sqrt{\frac{12}{c}} +1\right) }.
\label{k(c)}
\end{equation}
The scaling of the correlation length looks like
\begin{equation}
\xi_D\propto D^{k}
\label{xiscaling}
\end{equation}
again with  corrections of $O(1/\log D)$, see Ref. \cite{pollmann}, \cite{ujpirvu}, \cite{tagliacozzo} and \cite{andersson} for more details.

In practice however, one should go in the other way round: Testing the scaling relations one wants to decide whether the system is critical or not by giving an estimate for the correlation length.
The closer we are to the critical point the more delicate the problem due to the high correlation length.
One has to decide if the limits $S=S_{\infty}=\lim_{D\to\infty} S_D$ and $\xi=\xi_{\infty}=\lim_{D\to\infty} \xi_D$ are finite or not. ($S$ and $\xi$ converge to the exact values of the infinite system as the Trotter errors tend to zero.)
It can be especially hard for example in the vicinity of a Kosterlitz-Thouless transition when the inverse of the true correlation length $\xi^{-1}$ tends to zero very quickly as we are approaching the critical point.
As in finite-size scaling let us assume some scaling function for $\xi_D$ as
\begin{equation}
\log \xi_D=k \log D+g(D,\xi_{\infty})
\label{finitescaling}
\end{equation}

Clearly, $g(D,\xi_{\infty})\approx \log(\xi_{\infty}/D^k)$ for large $D$ if $\xi_D\to\xi_{\infty}<\infty$. 
It enables us to define $f(x)=g(D,\xi_{\infty})$ with $x=D^k/\xi_{\infty}$, for which $f(x)\to 0$ if $x\to 0$ and $f(x)\sim -\log(x)$
for large $x$.
Note that besides these limit properties we will not make any further observations about the form of $f$, just use a simple ansatz what will be justified by the numerics to be a good indicator for the non-critical behaviour.( See below the similar ansatz function for the finite-size case). 
So for instance, the following simple ansatz meets these requirements
\begin{equation}
f(x)=-\log(x+e^{-\alpha x})+const,\qquad \alpha>0.
\label{ansatz}
\end{equation}

Near the critical regime $\xi$ is big but not infinite and for a not big enough $D$ it behaves very similarly to (\ref{xiscaling}) and one cannot distinguish it from the critical scaling.
So the natural question arises: 
For a given and large but finite $\xi$ near the critical regime, at least how big $D$ should be applied in order to rule out the critical scaling with some reasonable accuracy? 
Essentially it depends on the ratio of $\log D$ and $f(\frac{D^{k}}{\xi_{\infty}})$.

In order to decide whether a given point is critical or not one should test the linearity of $\log \xi_D$ against $\log D$ as a null hypothesis. If it can be rejected then one can fit a trial function according to (\ref{finitescaling}) and (\ref{ansatz}) to obtain an estimate for $\xi_\infty$.

Note that(\ref{finitescaling}) resembles of the finite-size scaling of the entropy of the half chain of length $L$ applied in 
Ref. \cite{annni} 
\begin{equation}
S(L)=\frac{c}{6}(\log L+f(L\xi_{\infty}^{-1}))
\label{finitescaling2}
\end{equation}
with the same ansatz for $f(x)$ but now one has to plug $x=L/\xi_{\infty}$.
Note that this simple ansatz (\ref{ansatz}) for $f(x)$ has the right limits but for intermediate values it is not correct, however it turns out to be good approximation for $S(L)$ and a good indicator for non-criticality by the numerics.
One can obtain (universal) series expansions for $f(x)$ in the small and large limit theoretically, see Ref. \cite{cardy}. One can also confront this with our ansatz which incorporates the to limits.  

Of course here, for a given $L$, the applied bond dimension (or the number of states kept in each DMRG iteration) should be "high enough".

In our case however, we only have the bond dimension as an input parameter which simplifies the picture.
Besides, the main advantage is as follows:
Having the same $\xi_{\infty}$ for the infinite system, for the same accuracy one has considerably smaller computational cost if the scaling exponent $k>1$. (This will turn out to be the case in the present model as the central charges are found to be $c=1$ implying $k\approx 1.34$ and $c=1/2$ implies $k\approx 2.03$.)
In any case, the time cost of one update step scales as $O(LD^3)$ in the finite size case as opposed to $O(D^3)$ for the infinite case.
Now, if one takes the same maximal bond dimension $D$ for both cases, then for the finite-$D$ scaling one needs to also apply some smaller $D$-s which makes the overall time costs scale as $O(D^4)$ (at most, if one passes along $1,2...D-1,D$).
On the other hand in case of finite-size scaling the time costs $\propto O(L^2D^3)$.
In order to detect off critical scaling with the same accuracy one expects the necessary maximal length $L$ in the finite-size case must be bigger than the maximal $D$ in the finite-D case, if $k>1$. Nevertheless, this can be estimated by studying the scaling anzatz (\ref{finitescaling}) and (\ref{finitescaling2}).
In terms of $D$  this would result in an overall effective scaling  higher than $ O(D^5)$, in the finite-size case, obtained by the assumption $L\propto D$.

\section{Results}

\begin{figure}

\centering

\includegraphics[width=16cm, height=8cm]{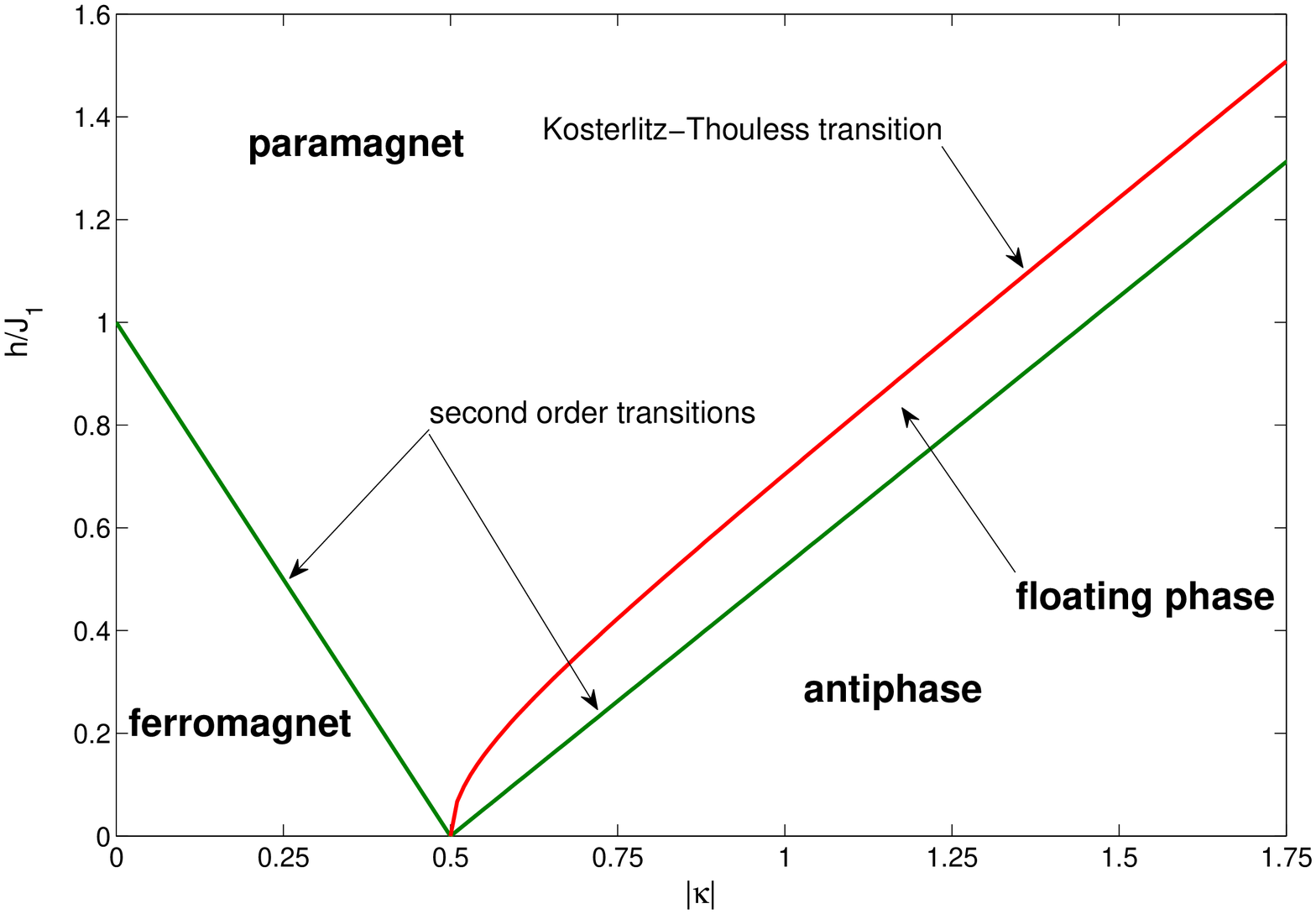}

\caption{ \label{fig:1} Schematic phase diagram}

\end{figure}

Figure [\ref{fig:1}] shows the phase diagram as it has been found in earlier studies by simulations Ref.\cite{annni} and perturbative analysis Ref.\cite{indiai}.
We try to explore the phase boundaries by looking at the correlation lengths as a function of the parameters and the second derivative of the energy. A Kosterlitz-Thouless type transition is conjectured between the floating and the paramagnetic(PM) phase where
the derivatives of the energy are useless indicators so the expected divergence of the correlation length is checked only.
In the critical points we also check the finite entanglement scaling laws and find very good agreement with the theoretical results. Moreover, in some special points we calculate the critical charges as well as the correlation scaling exponents.

We intend to determine some points of the line of phase transitions between the distinct phases and
check these relations as well as the finite-entanglement scaling and compute the central charge.
There are three lines of phase transitions, two of them are of second order while an infinite order transition is expected for the third one which needs a special care.
We also try to detect the floating phase when $\kappa<0$ and $|\kappa|$ is large, since its existence for arbitrary large $|\kappa|$ is still an open question.

\subsection{FM-PM transition}

In case of the Ising model, in the vicinity of the critical point as a function of the magnetic field, the correlation length 
and the magnetization diverges as
$$
\xi\propto |h-h_c|^{-\nu}, \qquad M\propto |h-h_c|^{\beta}
$$
and the energy and its first derivative are continuous but its second derivative diverges at $h=h_c$.
Using $D=40$ we get $0.535<h_c<0.538$ by looking at the extrema of the finite derivative $\partial^2E/\partial h^2$.
At the transition corresponding to $\kappa=-0.25$ we find the same behaviour for the energy as a function of $h$ and a very good agreement with $\nu=1$ as in the simple Ising case: $\nu=1.02$ comes from the linear fit on the figure below.
\begin{figure}[h!]

\centering

\includegraphics[width=18cm, height=8cm]{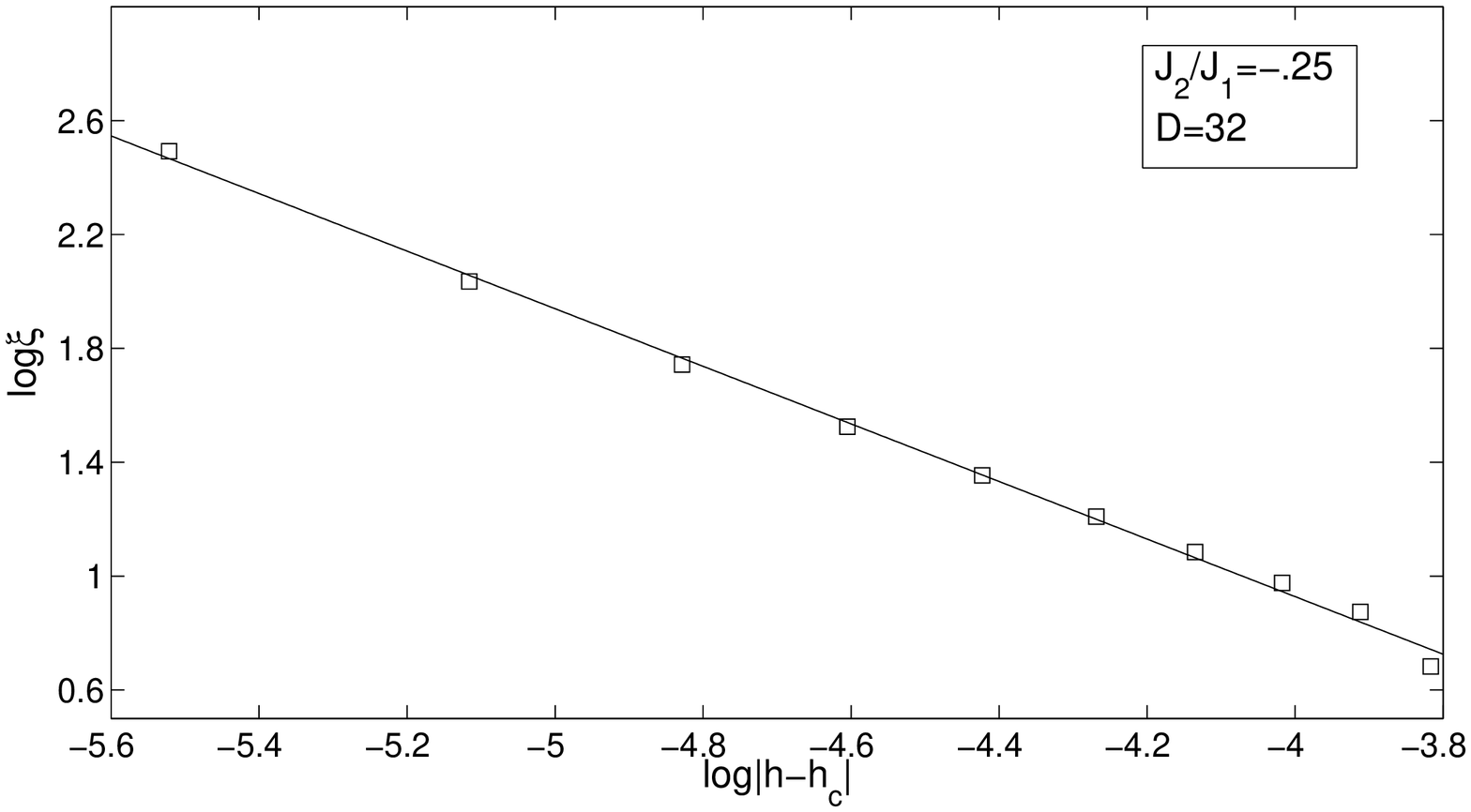}

\caption{Estimation of the scaling exponent at $\kappa=-0.25$ paramagnetic-ferromagnetic transition}\label{fig:2}

\end{figure}

\subsection{Antiphase-floating phase transition}

\begin{figure}[h!]

\includegraphics[width=10cm, height=6cm]{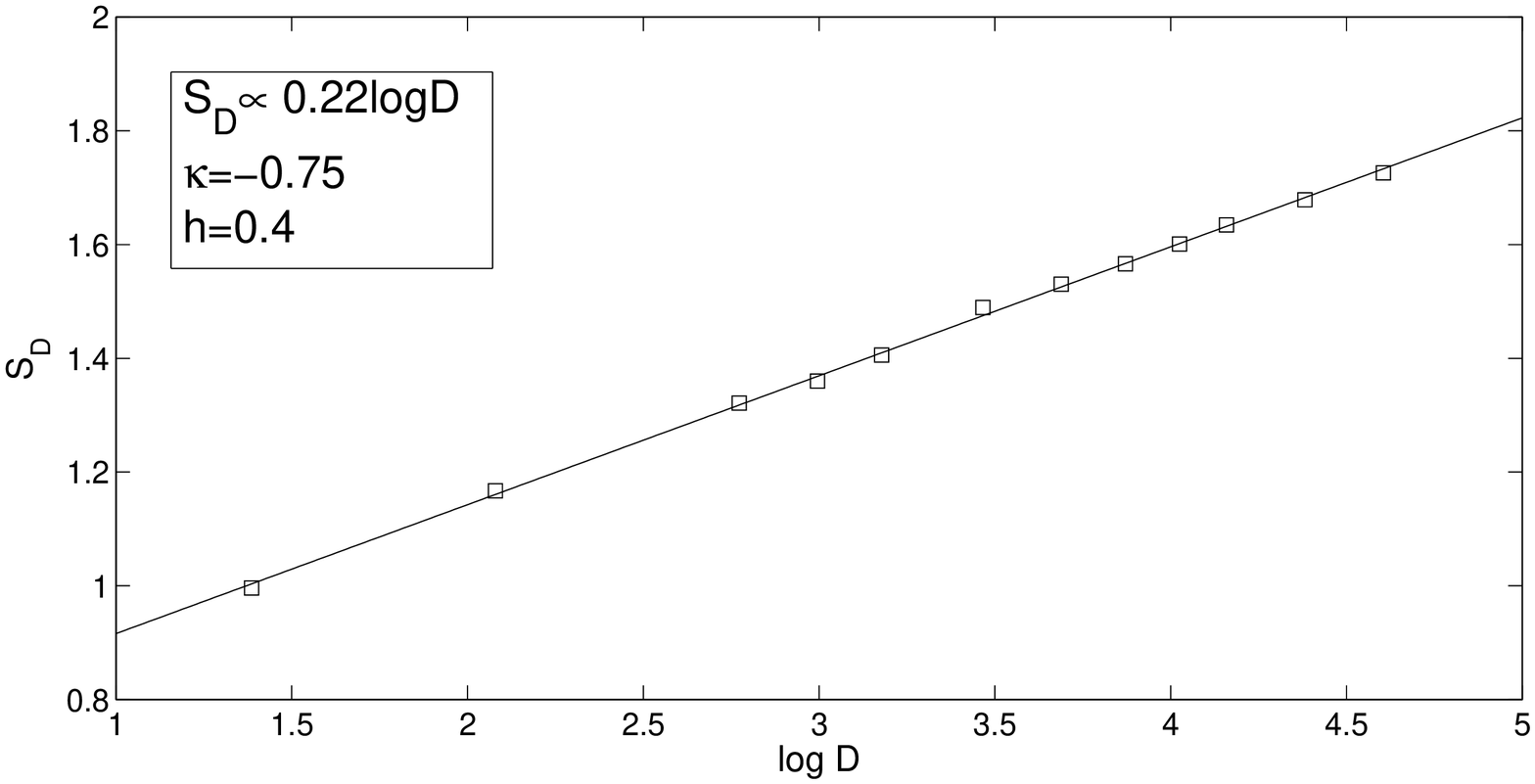}
\caption{Entanglement entropy scaling inside the floating phase at $\kappa=-0.75$, $h=0.4$ up to $D=100$.}
Central charge $c=1$ is suggested according to (\ref{entropyscaling}) and (\ref{k(c)}).
\label{fig:3}
\end{figure}

Again a second order phase transition is suggested by the numerics but with a central charge $c=1$ of the floating phase as opposed to the Ising case with $c=1/2$. See figure (\ref{fig:3}).

\subsection{ Floating phase-paramagnetic phase transition}

\begin{figure}[h!]

\centering

\includegraphics[width=15cm, height=9cm]{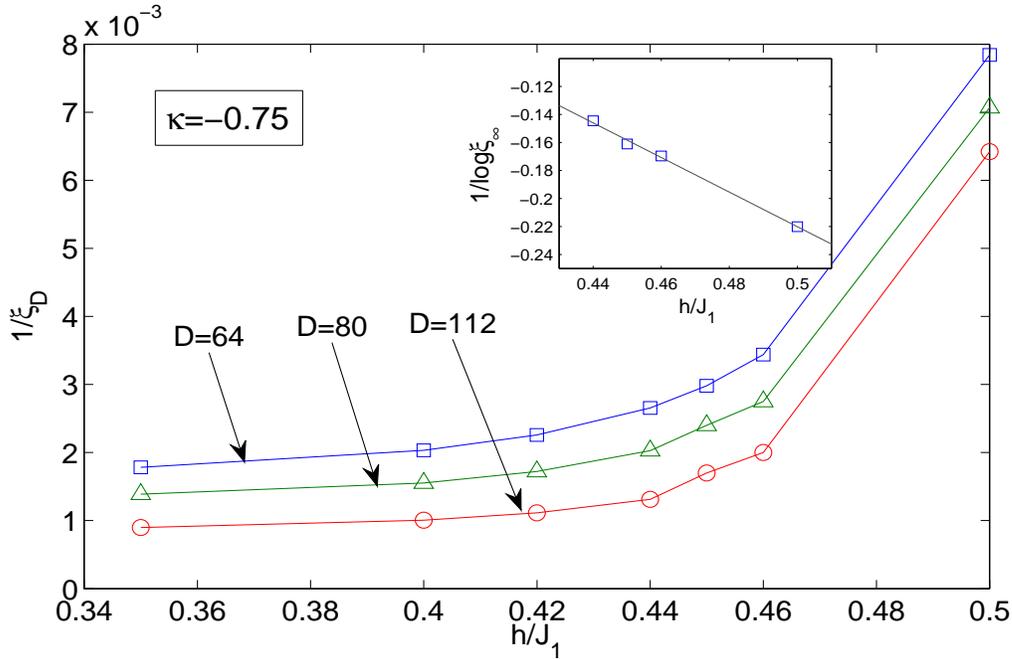}

\caption{Floating phase-paramagnetic phase transition illustrated with the inverse correlations plotted for three different bond dimensions. The inner plot shows $\1/\log(\xi_{\infty})$ vs $h$. Here, for a fixed $h$, $\xi_{\infty}$ is computed by fitting (\ref{finitescaling}). \label{fig:4} }

\end{figure}

\begin{figure}[h!]
\centering
\includegraphics[width=13cm, height=8cm]{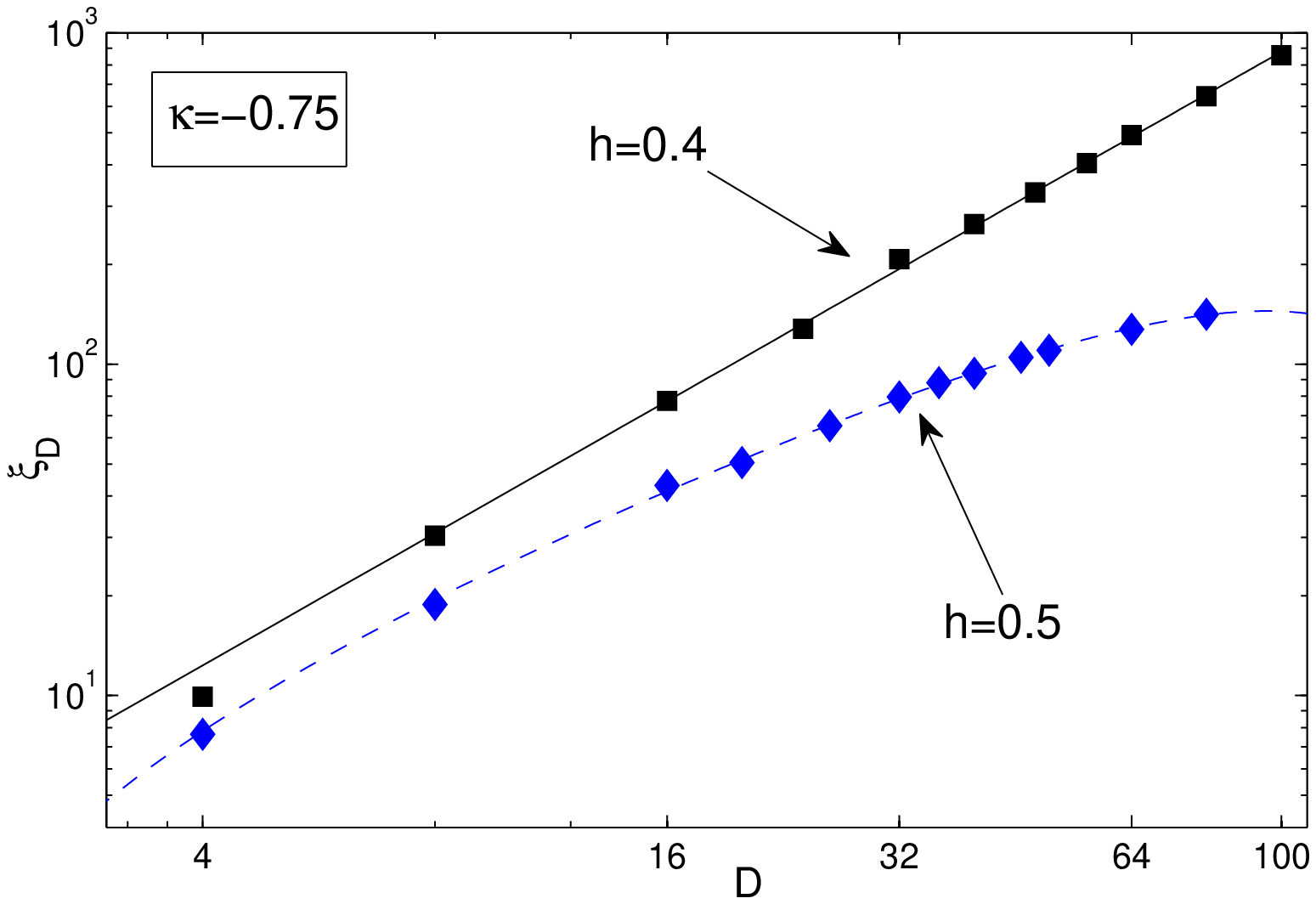}
\caption{$\xi_D$ against $D$ in the two sides of the floating-paramagnetic transition. It can be seen in the log-log scale how it bends down from the straight line for higher D, implying a non critical point at $h=0.5$. On the other hand, at $h=0.4$ the critical scaling hypotesis (\ref{xiscaling}) cannot be rejected with the applied maximal bond dimension}\label{fig:5}

\end{figure}
First, we check the power-law scaling of $\xi$ assumed for critical points as in (\ref{xiscaling}). If this power-law scaling   can be rejected statistically, then for a given $h$ and $\kappa$, $\xi_{\infty}$ is computed by the best fit using (\ref{finitescaling}) and finite-D scaling. (See the inset in FIG.\ref{fig:4}.)
We observe that  (\ref{finitescaling}) fits well using the ansatz function (\ref{ansatz}) with weak dependence of the parameter $\alpha$ which can be set to $1$.

In fact, the numerics indicate a higher order transition here, with continuous derivatives of the energy. Probably a Thouless-Kosterlitz type transition takes place as suggested by the plot $\xi^{-1}$ vs $h$ since $\xi(h)^{-1}\propto \exp(\frac{\alpha}{h_c-h})$ could be fitted if $h>h_c$, $\alpha>0$.
At the point $\kappa=-0.75$ around $h=0.45$ very close to the floating/PM transition we need about $D>80$ to point out the non power-law scaling (see. FIG. \ref{fig:5}). Otherwise, where we cannot reject the power-law scaling hypothesis, even though a finite value for $\xi_\infty$ obtained by fitting (\ref{finitescaling}) is rejected. In these cases one cannot rule out the critical scaling statistically. We have got anyway quite ambiguous results for $\xi_\infty$ which --in this case--can be regarded as a consequence of the errors of the simulation.
As in Ref. \cite{annni}, in FIG. \ref{fig:4} we plot the inverse correlation length along the line $\kappa=-0.75$ near to the supposed Kosterlitz-Thouless transition and find similar results with those calculated by finite DMRG and studying domain wall energies. Having our division for the magnetic field, the critical value can be estimated roughly as $0.42<h_c<0.44$. Note that here, the study of model specific quantities have been avoided.

\subsection { Large frustration}

\begin{figure}[h!]
\centering
\includegraphics[width=16cm, height=9cm]{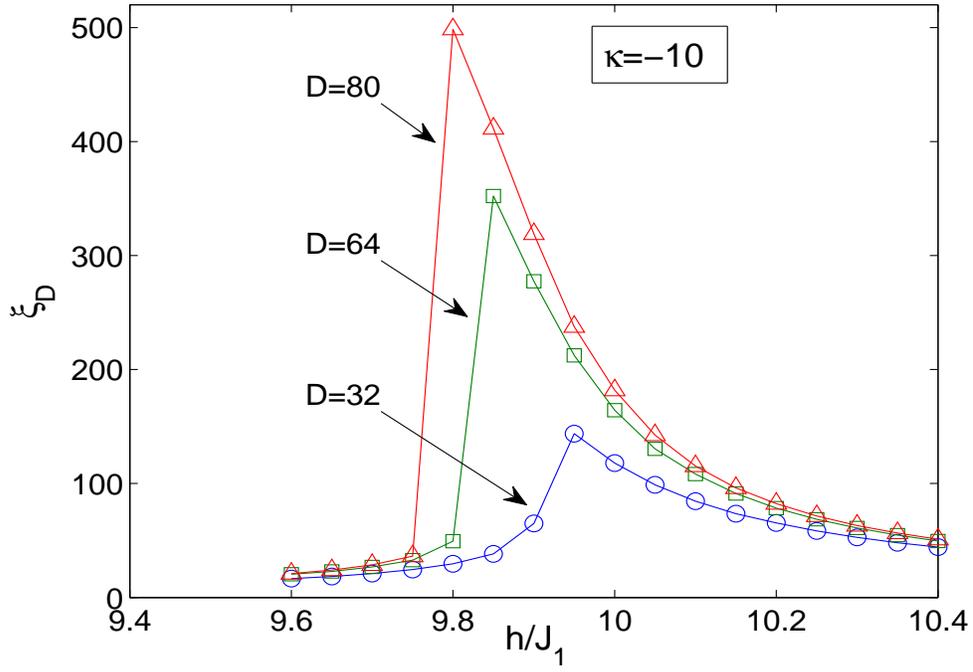}
\caption{Sharp peaks of the correlation length against the magnetic field using different bond dimensions at $\kappa=-10$ around the phase transition.}\label{fig:6}
\end{figure}

At $\kappa=-10$ up to the the applied maximal bond dimension $D_{max}=80$ and the division for the magnetic field, the numerics do not seem to indicate the presence of the floating phase because one sees a sharp peak plotting  $\xi$ against $h$. See Fig.\ref{fig:6}.

\begin{figure}[h!]
\centering
\includegraphics[width=16cm, height=9cm]{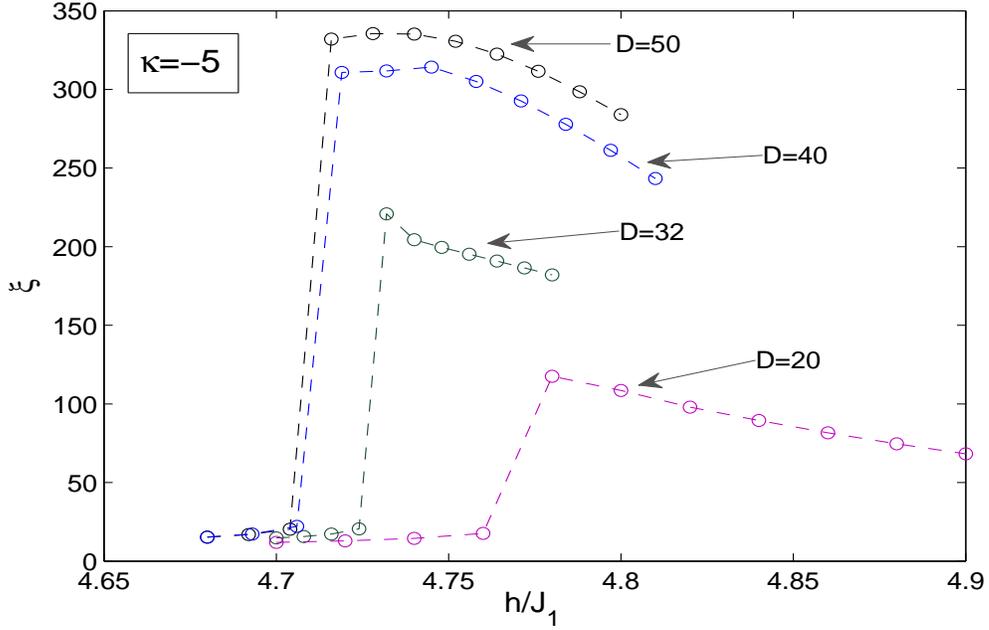}
\caption{The correlation length against the magnetic field using different bond dimensions at $\kappa=-5$ indicates an extended critical interval.}\label{fig:7}
\end{figure}

Note that in Ref. \cite{annni} the existence of the floating phase up to $\kappa=-5$ has been indicated but the DMRG method has become imprecise for higher frustrations, however with our new method, the floating phase can be suspected even with $D_{max}=50$ as shown in FIG. \ref{fig:7}. 

\section{Summary and conclusions}
The phase diagram of the 1D transverse ANNNI model is studied. We have coupled together neighboring sites of the chain in order to get NN interactions between the $4$-dimensional superspins and applied the simple iTEBD  with the new NN Hamiltonian.
In practice, the iTEBD algorithm seems to be more beneficial than others based on the transfer matrix and its leading eigenvectors. 
At some typical points of the phase diagram we have studied the phase transitions and estimated the critical quantities and found good agreement with previous DMRG results coming from finite-size analysis and model specific quantities.
We have also checked finite-entanglement scaling and its breaking to localize a supposedly infinite order transition between the floating and the paramagnetic phase, as well as pointed out the logarithmic divergence of the entanglement entropy inside the floating phase. With the proposed scaling ansatz for $\xi$ one can identify the critical regime with less computational effort in comparison with finite-size scaling, hence for the same results lower bond dimensions are sufficient.
For a high frustration parameter $\kappa=-10$ the numerics do not indicate the presence of the floating phase with the applied maximal bond dimension $D=80$ and resolution of the magnetic field $\Delta h=0.08$. However, we can suspect the existence of the floating phase for $\kappa=-5$ already with $D<=50$. The question of its existence for large frustrations requires a more detailed study with higher bond dimensions, but the applied methods throughout the paper are quite universal and applicable to other frustrated spin systems as well.

\section{Acknowledgment}
The author is grateful for the stimulating discussions with Lorenzo Campos Venuti and Marco Roncaglia, as well as their hospitality at the I.S.I. Foundation in Turin, Italy.

\end{document}